\documentclass[aps,pra,twocolumn,superscriptaddress,showkeys,amsmath,amssymb,longbibliography]{revtex4-1}		

\usepackage{graphicx}
\usepackage{dcolumn}
\usepackage{bm}
\usepackage[colorlinks,urlcolor=blue,citecolor=blue,linkcolor=blue]{hyperref}
\usepackage{subfig}
\usepackage[font=small,labelfont=bf,format=plain,justification=centerlast,labelsep=period]{caption}
\usepackage{tikz}
\usetikzlibrary{arrows,shapes,positioning,shadows,svg.path,backgrounds,fit}

\usepackage{verbatim}
\usepackage[normalem]{ulem}
\usepackage{float}
\usepackage{orcidlink}

\begin{document}

\title{Vacuum radiation versus shortcuts to adiabaticity}

\author{Ricardo R. Ancheyta\,\orcidlink{0000-0001-6718-8587}}
\email{ancheyta@fata.unam.mx}
\affiliation{Centro de F\'isica Aplicada y Tecnolog\'ia Avanzada, Universidad Nacional Autónoma de M\'exico, Boulevard Juriquilla 3001, Quer\'etaro 76230, Mexico}


\begin{abstract}
The nonadiabatic dynamic of the electromagnetic field triggers photons generation from the quantum vacuum. Shortcuts to adiabaticity, instead, are protocols that mimic the field's adiabatic dynamic in a finite time. Here, we show how the counterdiabatic term of the transitionless tracking algorithm cancels out, exactly, the term responsible for the photon production in the dynamical Casimir effect. This result suggests that the energy of producing photons out of the vacuum is related to the energetic cost of the shortcut. Furthermore, if the system operates under a quantum thermodynamic cycle, we confirm the equivalence between the adiabatic and nonadiabatic work outputs. Finally, our study reveals that identifying these unreported observations can only be possible using the so-called effective Hamiltonian approach.
\end{abstract}

\date{\today}

\maketitle

\section{Introduction}

From where there is no light (vacuum), photons can be created if the electromagnetic field is subject to nonadiabatic changes. This impressive phenomenon is known as the dynamical Casimir effect (DCE)~\cite{Dodonov_2020,Dodonov_2010}. It was first predicted using a variable-length cavity~\cite{Moore_1970} and experimentally confirmed within the architecture of superconducting quantum circuits~\cite{Wilson_Nature2011,Paraoanu_PNS_2013}. Besides being fundamental to quantum field theory~\cite{Nori_RMP_2012}, the DCE may have applications in nanophotonics, nanomechanics, and chemistry since it modifies the static Casimir force~\cite{Dodonov_2020,Palasantzas_2020,Munday_2021}. Moreover, the DCE is a finite-time quantum electrodynamics process and is expected to be relevant in finite-time quantum thermodynamics~\cite{Peter_Salamon_2022,Dodonov_JPA_2018,Quan_PRA_2019}. For example, in a quantum refrigerator at low temperatures, the DCE dominates and imposes the ultimate limit for cooling~\cite{Paz_PRE_2017}. The reason is a heating process produced by the nonadiabatic excitations of the DCE enforcing the third law~\cite{Benenti_PRA_2015}.

On the other hand, shortcuts to adiabaticity (STA) are protocols that imitate adiabatic systems' dynamics (quantum or classical) at finite times~\cite{Muga_RMP_2019,TORRONTEGUI2013117,Campo_IOP_2019}. STA are receiving much attention in quantum thermodynamics since they increase the power output of quantum heat engines by shortening their adiabatic (isentropic) strokes~\cite{Campo2014,Obinna_PRE_2018,Baris_PRE_2019,Obinna_PRE_2019,Obinna_PRR_2020}. STA act as quantum lubricants~\cite{Kosloff_PRE_2006}, producing frictionless evolution by counteracting nonadiabatic excitations. 
Theoretically, we may study the STA and the DCE with the harmonic oscillator. For instance, the work by Muga et al.~\cite{Muga_2010_JBP} introduced one of the first STA based on transitionless quantum driving~\cite{Berry_2009}; it changed an oscillator's frequency without transitions in finite time. For the DCE, one of its simplest versions occurs in a single-mode nonstationary cavity modeled by an oscillator with time-dependent frequency. 

In this work, using an effective Hamiltonian approach~\cite{Trembling_PRA_1998}, we describe in detail how the nonlocal potential arising from the STA combats the generation of photons in the DCE. We show that the squeezing terms appearing in the DCE and the STA are the same but with opposite signs, implying that the energy of producing photons from the vacuum relates to the energetic cost of the shortcut. Furthermore, with the effective Hamiltonian approach, we make transparent results of the STA in quantum thermodynamics, for instance, the equivalence between the adiabatic and nonadiabatic work outputs.

To make our article as self-consistent as possible, we structure it as follows. First, Sec.~\ref{TT_general} reviews the STA based on the transitionless tracking algorithm, and Sec.~\ref{TT_HO} discusses the STA of the quantum harmonic oscillator with time-dependent frequency. Next, Sec.~\ref{Sec3} describes the effective Hamiltonian approach of the DCE and shows that the squeezing term responsible for generating photons is the same (with the opposite sign) as the counterdiabatic term obtained in Sec.~\ref{TT_HO}. This observation simplifies the corresponding Hamiltonian structure and makes it easier to compute the work output when the system operates under a quantum thermodynamic cycle, Sec.~\ref{therm_section}. Finally, we show our conclusions in Sec.~\ref{Sec4}. 

\section{Shortcuts to adiabaticity}

\subsection{Transitionless tracking algorithm}\label{TT_general}

We review Berry’s formulation of the shortcuts to adiabaticity based on the transitionless tracking algorithm since it is more straightforward and enlightening~\cite{Berry_2009, Muga_RMP_2019}. For the time-dependent Hamiltonian $\hat{H}_0(t)$ satisfying $\hat{H}_0(t)|n(t)\rangle =E_n(t)|n(t)\rangle$, the adiabatic approximation states that $|\psi_n(t)\rangle=e^{{\rm i}\gamma_n(t)}e^{{\rm i}\theta_n(t)}|n(t)\rangle$
is an \textcolor{black}{{\it approximate}} solution of the Schr\"odinger equation when $\hat H_0(t)$ varies slowly, see~\textcolor{black}{Appendix~\ref{Appendix_1}.} Here, $\gamma_n(t)= {\rm i}\int_0^t dt'\langle n(t')|[{\partial_{t'}}|n(t')\rangle]$ and $\theta_n(t)=-\int_0^t dt' E_n(t')\hbar^{-1}$ are the geometrical and dynamical phases. The aim of the transitionless tracking algorithm is to find a quantum Hamiltonian $\hat H(t)$ such that $|\psi_n(t)\rangle$ is an \textcolor{black}{{\it exact}} solution of the corresponding Schr\"odinger equation ${\rm i}\hbar\,\partial_t|\psi_n(t)\rangle =\hat H(t)|\psi_n(t)\rangle$.
The time-evolution operator $\hat U(t)$ also satisfies the Schr\"odinger equation ${\rm i}\hbar\,\partial_t \hat U(t)=\hat H(t)\hat U(t)$,
then $\hat H(t)$ can be extracted from $\hat H(t)={\rm i}\hbar \,{\partial_t \hat U(t)}\,\hat U^\dagger(t)$.
The next step is to find an explicit form of the evolution operator. Fortunately, by choosing $\hat U(t)={\sum}_n |\psi_n(t)\rangle\langle n(0)|$,
we obtain $\hat U(t)|m(0)\rangle=|\psi_m(t)\rangle$. Therefore, $\hat U(t)$ is indeed an appropriate choice, where its time-derivative is
\begin{equation}
\begin{split}
    \frac{\partial \hat U(t)}{\partial t}=&{\sum}_n \Big\lbrace {\rm i}\,\dot \gamma_n(t)\, |\psi_n(t)\rangle\langle n(0)|
    \\
    &\qquad+{\rm i}\,\dot\theta_n(t)\,|\psi_n(t)\rangle\langle n(0)|
    \\
    &\qquad\quad +e^{{\rm i}\gamma_n(t)}e^{{\rm i}\theta_n(t)}\Big[\frac{\partial}{\partial t}|n(t)\rangle\Big]\langle n(0)|
    \Big\rbrace.
\end{split}
\end{equation}
Multiplying the above equation with $\hat U^\dagger(t)$ we get
\begin{equation}
    \begin{split}
         \frac{\partial \hat U(t)}{\partial t} &\hat U^\dagger (t)=
         {\sum}_n \Big\lbrace {\rm i}\,\dot\gamma_n(t)\, |n(t)\rangle\langle n(t)|
         \\
         &\quad+{\rm i}\,\dot\theta_n(t)\, |n(t)\rangle\langle n(t)|
         +\Big[\frac{\partial}{\partial t}|n(t)\rangle\Big]\langle n(t)|
         \Big\rbrace.
    \end{split}
\end{equation}
Therefore, the searched Hamiltonian is
\begin{equation}
    \hat H(t)={\sum}_n \big\lbrace 
        E_n|n\rangle\langle n|+{\rm i}\hbar (|\partial_t n\rangle\langle n|-\langle n|\partial_t n\rangle |n\rangle\langle n|)\big\rbrace.
\end{equation}
Note we have adopted the simplified notation $|n\rangle=|n(t)\rangle$, $E_n=E_n(t)$ and $|\partial_t n\rangle=\partial |n(t)\rangle/\partial t$. One can split $\hat H(t)$ in two parts, $\hat H(t)\equiv\hat H_0(t)+\hat H_1(t)$, where $\hat H_0(t)={\sum}_n E_n|n\rangle\langle n|$ and $\hat H_1(t)={\sum}_n {\rm i}\hbar\,(|\partial_t n\rangle\langle n|-\langle n|\partial_t n\rangle |n\rangle\langle n|)$
are known as the reference and counterdiabatic Hamiltonian, respectively~\cite{Berry_2009}.

\subsection{Time-dependent quantum harmonic oscillator}\label{TT_HO}

We choose the reference Hamiltonian $\hat H_0(t)$ as the quantum harmonic oscillator Hamiltonian with time-dependent frequency, $\omega(t)$ (with $m=1$)
\begin{equation}\label{H_0_QHO}
    \hat H_0(t)=\frac{\hat p^2}{2}+\frac{1}{2}\omega^2(t)\hat x^2,
\end{equation}
where $\hat x$ and $\hat p$ are, respectively, the position and momentum quantum mechanical operators satisfying the canonical commutation relation $[\hat x,\hat p]={\rm i}\hbar$. The corresponding instantaneous eigenvalues are $E_n(t)=\omega(t)(n+1/2)$. Note we did the mass of the oscillator to the unit to connect with the quantum Hamiltonian of the non-stationary electromagnetic field in Sec.~\ref{Sec3}. To get $\hat H_1(t)$ one needs to know the instantaneous eigenstates $|n(t)\rangle$ of $\hat H_0(t)$. In the coordinate representation, these are
\begin{equation}\label{eigenfunctions_n}
    \begin{split}
        \langle x|n(t)\rangle=\frac{1}{\sqrt{2^n n!}}\left[\frac{\omega(t)}{\pi\hbar}\right]^\frac{1}{4}
         H_n & \left(x\sqrt{{\omega(t)}/{\hbar}}\right) 
        \\
        & \times \exp\left(-\frac{\omega(t)}{2\hbar}x^2\right),
    \end{split} 
\end{equation}
where $H_n$ are the Hermite polynomials. Using the recurrence relation of $H_n$, Muga et al.~\cite{Muga_2010_JBP} proved that $\langle n|\partial_t n\rangle=0$, simplifying $\hat H_1(t)={\rm i}\hbar\sum_n |\partial_t n\rangle\langle n |$. By computing the time-derivative of (\ref{eigenfunctions_n}) one can show that the counterdiabatic term for the time-dependent quantum harmonic oscillator is~\cite{Muga_2010_JBP}
\begin{equation}\label{H1_counter_xp}
    \hat H_1(t)=-\frac{\dot\omega(t)}{4\,\omega(t)}(\hat x\hat p+\hat p\hat x),
\end{equation}
where the dot in $\omega(t)$ represents its time-derivative. The above result implies that $|\psi_n(t)\rangle$, see Sec.~\ref{TT_general}, is \textcolor{black}{an exact} solution of the Schr\"odinger equation when using
\begin{equation}\label{H_tot_xp}
\hat H(t)=\frac{\hat p^2}{2}+\frac{1}{2}\omega^2(t)\hat x^2-\frac{\dot\omega(t)}{4\,\omega(t)}(\hat x\hat p+\hat p\hat x).
\end{equation}
Hamiltonian (\ref{H1_counter_xp}) is a non-local operator, representing a non-local potential. Since it has a quadratic structure in terms of $\hat x$ and $\hat p$, Hamiltonian (\ref{H_tot_xp}) can be viewed as a generalized harmonic oscillator. Although the \textcolor{black}{finite time} dynamics generated by $\hat H(t)$ follows, exactly, the adiabatic \textcolor{black}{solution} of the reference Hamiltonian $\hat H_0(t)$, we want to emphasize that by only looking at the algebraic structure of Eq.~(\ref{H_tot_xp}), it is not obvious understanding why this happens. In other words, at this point, we do not know in detail how the new counterdiabatic term,  proportional to $\hat x\hat p+\hat p\hat x$, combats the nonadiabatic evolution generated by the Hamiltonian of the time-dependent harmonic oscillator~(\ref{H_0_QHO}). Clarifying such a question is, precisely, the aim of this work. 

First, note that $\hat H_1(t)$ in (\ref{H1_counter_xp}) is associated with the squeezing operator of quantum optics~\cite{gerry_knight_2004,agarwal2012quantum,klimov2009group}, as demonstrated in~\cite{Muga_2010_JBP,Moya_2003,Guasti_2003}. Second, it is well known that the time-dependent quantum harmonic oscillator generates, during its evolution, a squeezing effect in the system's quadratures. Therefore, it is reasonable to think that in (\ref{H_tot_xp}) the counterdiabatic term $\hat H_1(t)$ should somehow compensate for the squeezing evolution generated by $\hat H_0(t)$.

Due to the above discussion, we want to write (\ref{H_tot_xp}) in terms of ladder operators. However, writing the proper creation and annihilation operators of the time-dependent quantum harmonic oscillator, equivalent to a time-dependent spacetime metric, is by no means a trivial task~\cite{Lewis_1967, MALKIN1970243, Guasti_2003, Ibanez_PRL_2012,Natalia_PRD_2022,Natalia_PRA_2022}. We, for instance, may work with the ``instantaneous'' ladder operators given by~\cite{Muga_2010_JBP} ($\hbar=1$)
\begin{equation}
\begin{split}\label{inst_ladder}
\hat a_t^{}=\frac{1}{\sqrt{2\omega(t)}}\left[\omega(t)\,\hat x+{\rm i}\,\hat p\right],
\quad
\hat a^\dagger_t=\frac{1}{\sqrt{2\omega(t)}}[\omega(t)\,\hat x-{\rm i}\,\hat p].
\end{split}
\end{equation}
Evidently, for any given protocol that starts at $t=t_0$ and ends at $t=t_f^{}$, the oscillator's frequency $\omega(t)$ has fixed values $\omega_0^{}$ and $\omega_f^{}$ for $t\leq t_0$ and $t\geq t_f^{}$, respectively. For those cases, Eq.~(\ref{inst_ladder}) reduces to the well-known definition of the ladder operators of the harmonic oscillator with constant frequency~\cite{Sakurai}. Equation (\ref{inst_ladder}) implies that $\hat x=(\hat a^\dagger_t+\hat a_t^{})/{\sqrt{2\omega(t)}}$ and $\hat p={\rm i}(\hat a^\dagger_t-\hat a_t^{})\sqrt{\omega(t)/2}$, which upon substitution in (\ref{H_tot_xp}) one gets~\cite{Muga_2010_JBP}
\begin{equation}\label{H_tot_adda}
    \hat H(t)=\omega(t)\Big(\hat a^\dagger_t \hat a_t+\frac{1}{2}\Big)-{\rm i}\frac{\dot\omega(t)}{4\,\omega(t)}\big(\hat a^{\dagger 2}_t-\hat a_t^2\big).
\end{equation}
Again, the above equation does not provide us with a clear view of how it generates the adiabatic dynamics of reference Hamiltonian $\hat H_0(t)$. Instead, it brings us an important result, the second (squeezing) term of (\ref{H_tot_adda}) can be written without the temporal subscript $t$~\cite{Muga_2010_JBP}. This can be done due to the fact that for any frequency modulation $\omega(t)$, the combination ${\rm i}\hat a^{\dagger 2}_t-{\rm i}\hat a_t^2$ is always equal to the time-independent term $\hat x\hat p+\hat p\hat x$, thus, the former can be safely computed at any given time. In particular, and for convenience of the discussion of Sec.~\ref{Sec3}, we evaluate it at the beginning of a given time-dependent protocol, i.e., throughout the article we shall always write $\hat H_1(t)=-{\rm i}\dot\omega(t)(\hat a^{\dagger 2}_0-\hat a_0^2)/4\omega(t)$. On the contrary, the first term of (\ref{H_tot_adda}), $\hat a^{\dagger}_t \hat a_t+1/2$, does not equal any time-independent combination of $\hat x$ and $\hat p$. Actually, the reference Hamiltonian (\ref{H_0_QHO}) written in terms of the instantaneous ladder operators (\ref{inst_ladder}) at $t=t_0$ looks like
\begin{equation}\label{H_0_adda_0}
\begin{split}
    \hat H_0(t)=\omega_0^{}\Big(\hat a^\dagger_0\hat a_0^{}+\frac{1}{2}&\Big)\Big[\frac{\omega^2(t)}{2\,\omega_0^2}+\frac{1}{2}\Big]
    \\
    &+\omega_0^{}\big(\hat a^{\dagger 2}_0+\hat a_0^2\big) \Big[\frac{\omega^2(t)}{4\,\omega_0^2}-\frac{1}{4}\Big].
\end{split}
\end{equation}
To avoid any confusion, keep in mind that $\omega(t)$ in above comes from the bare Hamiltonian (\ref{H_0_QHO}), while $\omega_0^{}$ is the frequency at which the instantaneous ladder operators (\ref{inst_ladder}) are defined. As expected, if $\omega(t)\rightarrow\omega_0$, $\hat H_1(t)=0$ and $\hat H(t)\rightarrow\hat H_0(t)=\omega_0(\hat a^\dagger_0\hat a_0^{}+1/2)$. When the frequency modulation is of the form $\omega(t)=\omega_0+f(t)$, with $f(t)$ an arbitrary (usually periodic) time-dependent function, $\hat H_0(t)=\omega_0(\hat a^\dagger_0\hat a_0^{}+1/2)+f^2(t)(\hat a^{\dagger 2}_0+\hat a_0^2+2\hat a^\dagger_0\hat a_0^{}+1)/4\omega_0$ has an algebraic structure that has been used to investigate the evolution of coherent states in a Kerr medium~\cite{Roman-Ancheyta:15}. It is worth mentioning that (\ref{H_0_adda_0}) possesses a time-independent squeezing term $(\hat a^{\dagger 2}_0+\hat a_0^2)$, which, unfortunately, cannot be eliminated with the one contained in $\hat H_1(t)$. However, in the next section, we explicitly show how it is possible to obtain an effective reference Hamiltonian $\hat H_0(t)$ containing the squeezing term that cancels out $\hat H_1(t)$. This idea naturally comes from the studies of non-stationary cavity fields in which the dynamical Casimir effect occurs.

\section{Dynamical Casimir effect}

\subsection{Effective Hamiltonian approach}\label{Sec3}

The DCE, a name introduced by Schwinger~\cite{Schwinger}, was predicted by Moore in 1970~\cite{Moore_1970}. However, the original theory did not have a Hamiltonian, so for many years the DCE was studied in the Heisenberg representation. It was not until 1994 that C. K. Law derived an effective Hamiltonian of the DCE, allowing an equivalent Schr\"odinger description~\cite{Law_1994}. Essential features of the DCE, as the generation of photons from the vacuum, are captured by the single-mode version of such effective Hamiltonian, which is the Hamiltonian of the time-dependent harmonic oscillator~\cite{Dodonov_2010,Dodonov_2020}. In what follows, we show the basic steps to obtain the single-mode effective Hamiltonian of the time-dependent oscillator.

First, we need the Heisenberg representation for the instantaneous ladder operators. We replace in (\ref{inst_ladder}) the Heisenberg representation of the position and momentum operators, i.e., $\hat a_{\texttt H}^{}\equiv[\omega(t)\,\hat x_{\texttt H}^{}+{\rm i}\,\hat p_{\texttt H}^{}]/\sqrt{2\omega(t)}$, where the subscript $\texttt H$ in $\hat x_{\texttt H}^{}$ and $\hat p_{\texttt H}^{}$ indicates these are in the Heisenberg picture. Second, we take the time-derivative
\begin{eqnarray}
        \frac{d\hat a_{\texttt H}^{}}{dt}
        &=&\frac{1}{\sqrt{2\omega(t)}}\Big[\omega(t)\frac{d{\hat x}_{\texttt H}^{}}{dt}+{\rm i}\frac{d\hat p_{\texttt H}^{}}{dt}\Big]
        \nonumber\\
        &&\hspace{1.5cm}+\frac{1}{\sqrt{2\omega(t)}}\frac{d\omega(t)}{dt}\hat x_{\texttt H}^{}
        \nonumber\\
        &&\hspace{2.4cm}-\frac{1}{[2\omega(t)]^{3/2}}\frac{d\omega(t)}{dt}\big[\omega(t)\hat x_{\texttt H}^{}+{\rm i}\,\hat p_{\texttt{H}}^{}\big],
        \nonumber\\
      &=&\frac{1}{\sqrt{2\omega(t)}}[\omega(t)\,\hat p_{\texttt H}^{}-{\rm i}\,\omega^2(t)\,\hat x_{\texttt H}^{}]
        \nonumber\\
      &&\hspace{2.4cm}+\frac{1}{[2\omega(t)]^{3/2}}\frac{d\omega(t)}{dt}\big[\omega(t)\hat{x}_{\texttt H}^{}-{\rm i}\,\hat p_{\texttt H}^{}\big],
      \nonumber\\
      &=&-{\rm i}\,\omega(t)\hat{a}_{\texttt H}^{}+\frac{1}{2\omega(t)}\frac{d\omega(t)}{dt}\hat a^\dagger_{\texttt H}.
      \label{eq_motion_a}
\end{eqnarray}
In the same manner,
\begin{equation}\label{eq_motion_add}
     \frac{d\hat a_{\texttt H}^{\dagger}}{dt}=+{\rm i}\,\omega(t)\hat{a}_{\texttt H}^{\dagger}+\frac{1}{2\omega(t)}\frac{d\omega(t)}{dt}\hat a_{\texttt H}.
\end{equation}
To get (\ref{eq_motion_a}) and (\ref{eq_motion_add}) we have substituted the corresponding equations of motion for the position and momentum operators $d\hat x_{\texttt H}^{}/dt=\hat p_{\texttt H}^{}$ and $d\hat p_{\texttt H}^{}/dt=-\omega^2(t)\hat x_{\texttt H}^{}$. At this point, we stress that Eqs.~(\ref{eq_motion_a}) and (\ref{eq_motion_add}) {are not} proper quantum mechanical equations of motion~\cite{Sakurai}. Instead, we obtained them by deriving with respect to time the definition of the instantaneous ladder operator. However, we did use the proper equations of motion of the position and momentum operators generated by the Hamiltonian of the time-dependent harmonic oscillator, i.e., $\hat H_0(t)$ of (\ref{H_0_QHO}) in the Heisenberg equation $d \hat O_{\texttt H}^{}/dt={\rm i }[\hat H_0(t), \hat O_{\texttt H}^{}]$, where $\hat O_{\texttt H}^{}$ is an arbitrary operator in the Heisenberg representation. Note that contrary to the constant frequency case, Eqs.~(\ref{eq_motion_a}) and (\ref{eq_motion_add}) display an extra term due to the explicit time-dependency of the ladder operators (\ref{inst_ladder}). 

Second,  to understand the idea of how the effective Hamiltonian approach works, we illustrate it with the constant frequency case. If $\omega(t)\rightarrow\omega_0^{}$, ${d\hat a_{\texttt H}^{}}/{dt}=-{\rm i}\,\omega_0^{}\hat{a}_{\texttt H}^{}$ and ${d\hat a_{\texttt H}^{\dagger}}/{dt}={\rm i}\,\omega_0^{}\hat{a}_{\texttt H}^{\dagger}$. Assuming we are, for some reason, only provided with the information on these equations rather than with the Hamiltonian itself. The crucial question is, which Hamiltonian in the Schr\"odinger picture can generate the equations above? If such a Hamiltonian exists, the provided equations may be interpreted as proper quantum mechanical equations of motion. For this simple case, it is easy to see that $\hat H_{\rm eff}=\omega_0^{}\hat a^\dagger_0\hat a_0^{}$ does the desired task since $[\hat H_{\rm eff},\hat a_0^{}]=-\omega_0^{}\hat a_0^{}$.

Now, for an arbitrary frequency drive $\omega(t)$ and noting that $[\hat a^{\dagger 2}_0,\hat a_0^{}]=-2\textcolor{black}{\hat a_0^\dagger}$, it is not difficult to show that the effective Hamiltonian,
\begin{equation}\label{H_eff}
    \hat H_{\rm eff}(t)=\omega(t)\hat a^\dagger_0\hat a_0^{}+{\rm i}\frac{1}{4\omega(t)}\frac{d\omega(t)}{dt}\big(\hat a^{\dagger 2}_0-\hat a_0^2\big),
\end{equation}
indeed generates Eqs.~(\ref{eq_motion_a}) and (\ref{eq_motion_add})~\cite{Dodonov_2020}. This effective Hamiltonian first derived almost three decades ago in the context of the DCE, describes an electromagnetic cavity with a moving mirror, where its frequency is given by $\omega(t)=\pi/q(t)$ and $q(t)$ is the prescribed mirror's trajectory~\cite{Law_1994}. The Hamiltonian of ~\cite{Law_1994} displays additional terms accounting for an intermode interaction induced by the non-stationary field's boundary conditions. Nevertheless, subsequent studies~\cite{Dodonov_PRA_2022,Ancheyta_2018,Roman_2017,Ancheyta_JOSAB_17,Dodonov_2013,PhysRevA.85.015805}  show that when the non-stationary cavity field supports one-single mode, (\ref{H_eff}) can be safely considered as the simplest version where the DCE can be manifested. For example, under the resonant conditions $\omega(t)=\omega_0[1+\varepsilon\sin (2\omega_0^{}t)]$, (\ref{H_eff}) predicts an exponential photon growth~\cite{Dodonov_2013,Roman_2017,Ancheyta_JOSAB_17} 
\begin{equation}\label{phton_growth}
    \langle 0|\hat U^\dagger\hat a^\dagger_0\hat a_0^{}\hat U|0\rangle=\sinh^2\left(\varepsilon\omega_0^{}t/2\right),
\end{equation}
where $\hat U$ is the corresponding time-evolution operator of $\hat H_{\rm eff}(t)$, $\hat a_0^{}|0\rangle=0$ defines the vacuum state, and  $\varepsilon\ll 1$ is a small amplitude modulation depth. Equation (\ref{phton_growth}) is known as the Casimir or vacuum radiation and is the consequence of the nonadiabatic boundary conditions of the field represented by the second term of~(\ref{H_eff}). 

Remarkably, when replacing $\hat H_0(t)$ by its physically equivalent effective Hamiltonian $\hat H_{\rm eff}(t)$ in the definition of $\hat H(t)$\textcolor{black}{, i.e., $\hat H(t)=\hat{H}_{\rm eff}(t)+\hat{H}_1(t)$, we obtain}
\begin{equation}\label{final_result}
    \hat H(t)=\omega(t)\,\hat a^\dagger_0\hat a_0^{}.
\end{equation}
This equation shows, crystal clear, that the finite-time dynamics generated by the shortcut to adiabaticity is, indeed, the one given by the quantum harmonic oscillator with an instantaneous frequency $\omega(t)$. Since the counterdiabatic term $\hat H_1(t)=-{\rm i}\dot\omega(t)(\hat a^{\dagger 2}_0-\hat a_0^2)/4\omega(t)$ cancels out the induced nonadiabatic squeezing term in~(\ref{H_eff}), the time evolution operator is $\hat U=\exp[-{\rm i}\int dt' \omega(t')\hat a^\dagger_0\hat a_0^{}]$, implying  $\langle0|\hat U^\dagger\hat a^\dagger_0\hat a_0^{}\hat U|0\rangle=0$. Therefore, no matter what frequency drive is used, photons generation from the vacuum is impossible when performing a shortcut to adiabaticity. Notice that from a different approach, a similar conclusion was recently obtained by exploiting the conformal symmetry of the system~\cite{Lombardo_PRA_2022}. 

Using (\ref{final_result}) it is easy to write the final energy, $\langle \hat H(t_f)\rangle=\omega_f^{}\langle \hat a^\dagger_0\hat a_0^{}\rangle$, in terms of the initial energy $\langle \hat H(t_0)\rangle$ as 
\begin{equation}\label{Energy_relation}
\langle \hat H(t_f)\rangle=(\omega_f/\omega_0^{})\langle \hat H(t_0)\rangle.
\end{equation}
The above expression is a common result \textcolor{black}{encountered in} the adiabatic limit \textcolor{black}{of systems with scale-invariance dynamical symmetry}, where $(\omega_0^{}/\omega_f^{})^{1/2}$ is known as the adiabatic scaling factor~\cite{Jaramillo_2016,delCampo2018}. However, since (\ref{final_result}) contains the use of an STA, \textcolor{black}{i.e., $\hat H(t)=\hat{H}_{\rm eff}(t)+\hat{H}_1(t)$,} the relation between initial and final energies is valid for any physically unitary finite-time protocol \textcolor{black}{driving $\hat{H}_{\rm eff}(t)$ through $\omega(t)$} and not just for the adiabatic approximation. \textcolor{black}{Certainly, $\hat H(t)$  strictly follows the adiabatic solution of the reference Hamiltonian $\hat{H}_0(t)$ but at a finite time.} An immediate consequence of this result is in finite-time quantum thermodynamics~\cite{Peter_Salamon_2022}. For instance, we may choose $\hat{H}(t)$ as the time-dependent Hamiltonian associated with the working substance of a reciprocating quantum heat engine~\cite{Obinna_PRL_2012, Robnagel_2016}. When the nonstationary cavity suffers an expansion (compression), the cavity length increases (decreases), and the frequency decreases (increases), ending with a lower (higher) internal energy (\ref{Energy_relation}). This behavior resembles the one encounter with a thermodynamic piston. Note that the previously discussed situation is not evident when using the standard expressions of $\hat H(t)$ encounter in (\ref{H_tot_xp}) or (\ref{H_tot_adda}). 

Even though the counterdiabatic term $\hat {H}_1(t)$ typically vanishes at the start ($t_0$) and end ($t_f$) of an STA, getting a relationship between initial and final energies using (\ref{H_tot_adda}) is not straightforward. This is because the initial (final) $\hat{H}(t)$ depends on the instantaneous operators at $t=t_0^{}\,(t_f)$. In contrast, $\hat{H}(t)$ in (\ref{final_result}) is always written in terms of operators at some specific time, we used $t_0$ but the result holds for other times.

\subsection{Quantum thermodynamic implications}\label{therm_section}

The quantum Otto cycle operates between a hot and a cold reservoir and is a paradigmatic thermodynamic cycle extensively used by the quantum thermodynamics community~\cite{Quan_PRA_2007, Kosloff_Otto}. The cycle consists of four branches: two adiabatic (isentropic) strokes where the working substance (the nonstationary cavity in our case) suffers a compression and expansion while isolated from the heat reservoirs;  two quantum isochoric strokes where the system is put in contact with one reservoir, here the cavity frequency has fixed values and heat transfer occurs (thermalization) but no work is performed. During the compression stroke, the cavity frequency, $\omega(t)$, increases from the initial value $\omega_1$ to $\omega_2$, while in the expansion, it goes from $\omega_2$ to $\omega_1$. At the hot (cold) isochoric stroke, the cavity relaxes to a thermal state with temperature $T_{\rm h}$ ($T_{\rm c}$). 

For the thermal state $\hat\rho_{\rm th}=\exp(-\hbar\omega_j\hat a^\dagger_0\hat a_0^{}/k_B^{}T)Z^{-1}$, where $Z$ is the partition function, the average number of photons is $\langle \hat a^\dagger_0\hat a_0^{}\rangle={\rm tr}\{\hat\rho_{\rm th}\hat a^\dagger_0\hat a_0^{}\}=\frac{1}{2}\coth(\hbar\omega_j/2k_B^{}T)-1/2$. With this and with the help of (\ref{Energy_relation}), we compute the average energies $\langle \hat H (t)\rangle$ of the nonstationary cavity at each of the four strokes of the Otto cycle. Since we use an STA, these values coincide with the ones obtained during the slow adiabatic process~\cite{Obinna_PRL_2012}. 
The mean work $\langle W\rangle$, calculated as ${\rm tr}\{\hat\rho\Delta \hat H\}$ with $\Delta \hat H$ the change in the Hamiltonian~\cite{Baris_PRE_2019}, during the compression and expansion strokes is $\langle W\rangle_{\texttt{comp}}=\frac{\hbar}{2}(\omega_2-\omega_1)\coth(\hbar\omega_1/2k_B^{}T_{\rm c})$ and  $\langle W\rangle_{\texttt{exp}}=\frac{\hbar}{2}(\omega_1-\omega_2)\coth(\hbar\omega_2/2k_B^{}T_{\rm h})$, respectively. Then, the total work per (finite-time) cycle, $\langle W\rangle_{\texttt{comp}}$+$\langle W\rangle_{\texttt{exp}}$, equals the work output in the slow adiabatic cycle. 
This expected result confirms that we have used an STA in the thermodynamic cycle. When computing the corresponding engine's efficiency, without the cost of the STA, one gets the maximum possible value $\eta=1-\omega_1/\omega_2$ at finite power~\cite{Campo2014}. Of course, to be fair with the slow adiabatic cycle, we still need to consider the thermodynamic cost of implementing the shortcut~\cite{Campo2014,Obinna_PRE_2018,Baris_PRE_2019,Obinna_PRE_2019,Obinna_PRR_2020,Paternostro_PRR_2023}. 
However, there is still an ongoing debate on how to compute this cost correctly and, just as important, how to incorporate it into the engine's efficiency, see~\cite{Muga_RMP_2019}.  Interestingly, any proposal willing to use the counterdiabatic term $\hat{H}_1(t)$ to compute the cost of the STA will tell us that the energetic cost of producing photons out of the vacuum is related to the cost of the shortcut since $\hat{H}_1(t)$ is the same as the squeezing term of the dynamical Casimir effect. 


\section{Conclusions}\label{Sec4}

We show that with the effective Hamiltonian approach of quantum optics, one can quickly identify that the counterdiabatic term of a shortcut to adiabaticity associated with the time-dependent quantum harmonic oscillator equals (with the opposite sign) the squeezing term inherent to the dynamical Casimir effect. We exhibit the combat between the nonadiabatic evolution generated by the DCE and the STA, the latter trying to enforce an adiabatic dynamic at all times. We confirm the equivalence between the adiabatic and nonadiabatic work outputs when using an STA in a quantum thermodynamic cycle. We also show that the energy cost of producing photons from the vacuum can be related to the shortcut cost. As we mainly deal with ladder operators, these results may be difficult to obtain only with the position and momentum operators, predominately used in the STA literature. Finally, it would be interesting to see other counterdiabatic Hamiltonians, believed to be challenging to implement, and look for their effective Hamiltonian counterpart and vice versa. 

\acknowledgments
The author acknowledges Bar{\i}\c{s} \c{C}akmak for indicating that the counterdiabatic Hamiltonian of the time-dependent harmonic oscillator is a squeezing term.

\appendix
\section{Adiabatic approximation}\label{Appendix_1}

For pedagogical purposes, here we follow the procedure carried out in~\cite{Sakurai} describing the adiabatic approximation in quantum mechanics. One starts with the Schr\"odinger equation ${\rm i}\hbar\,{\partial_t}|\psi(t)\rangle=\hat H_0(t)|\psi(t)\rangle$.
A general solution is $|\psi(t)\rangle=\sum_n c_n(t)e^{{\rm i}\theta_n(t)}|n(t) \rangle$, where $\theta_n(t)=-\frac{1}{\hbar}\int_0^t dt' E_n(t')$ is known as the dynamical phase, and $|n(t)\rangle$ satisfies $\hat H_0(t)|n(t)\rangle =E_n(t)|n(t)\rangle$. $c_n(t)$ are time-dependent expansion coefficients to be determined. Taking the time derivative of $|\psi(t)\rangle$ we get
\begin{equation}
\begin{split}
    \frac{\partial}{\partial t}|\psi(t)\rangle &= {\sum}_n\dot c_n(t)e^{{\rm i}\theta_n(t)}|n(t)\rangle
    \\
    &\qquad+{\sum}_n {\rm i}\,\dot\theta_n(t)c_n(t)e^{{\rm i}\theta_n(t)}|n(t)\rangle   
    \\
    &\qquad\qquad+{\sum}_n c_n(t)e^{{\rm i}\theta_n(t)}\frac{\partial}{\partial t}|n(t)\rangle
    \\
    &=\frac{1}{{\rm i}\hbar}\hat H_0(t)|n(t)\rangle.
\end{split}
\end{equation}
Using the eigenvalue equation and noting that $\dot\theta_n(t)=-E_n(t)/\hbar$, the above equation simplifies to
$\sum_n e^{{\rm i}\theta_n(t)}[\dot c_n(t)|n(t)\rangle+c_n(t){\partial_t}|n(t)\rangle]=0$. Taking the inner product with $\langle m(t)|$ and using the orthonormality associated with the eigenstates at equal times, one obtains the following differential equation
\begin{equation}
    \dot c_m(t)+{\sum}_n e^{{\rm i}[\theta_n(t)-\theta_m(t)]}\langle m(t)|\Big[\frac{\partial}{\partial t}|n(t)\rangle\Big]=0.
\end{equation}
On the other hand, by taking the time derivative for the eigenvalue equation we get $\dot{\hat H}_0(t)|n(t)\rangle+\hat H_0(t){\partial_t}|n(t)\rangle=\dot E_n(t)|n(t)\rangle+E_n(t){\partial_t}|n(t)\rangle$. It implies that
\begin{equation}
    \langle m(t)|\dot{\hat H}_0(t)|n(t)\rangle=[E_n(t)-E_m(t)]\langle m(t)|\Big[\frac{\partial}{\partial t}|n(t)\rangle\Big].
\end{equation}
This means that
\begin{equation}
    \langle m(t)|\Big[\frac{\partial}{\partial t}|n(t)\rangle\Big]=\frac{\langle m(t)|\dot{\hat H}_0(t)|n(t)\rangle}{E_n(t)-E_m(t)}.
\end{equation}
Substituting this in the equation for $\dot c_m(t)$ we get
\begin{equation}
\begin{split}
    \dot c_m(t)=&-c_m(t)\langle m(t)|\Big[\frac{\partial}{\partial t}|m(t)\rangle\Big]
    \\ & \qquad 
    -\sum_{n\neq m} c_n(t)e^{{\rm i}[\theta_n(t)-\theta_m(t)]}\frac{\langle m(t)|\dot{\hat H}(t)|n(t)\rangle}{E_n(t)-E_m(t)}.
\end{split}
\end{equation}
The adiabatic approximation consists in neglecting the second term in the right-hand side of the above equation~\cite{Sakurai}, 
$\dot c_m(t)\approx-c_m(t)\,\langle m(t)|\big[{\partial_t}|m(t)\rangle\big]$. This has the following solution
\begin{equation}
    c_m(t)=\exp\left(-\int_0^t dt'\langle m(t')|\Big[\frac{\partial}{\partial t'}|m(t')\rangle\Big]\right)c_m(0),
\end{equation}
which we can rewrite as $c_m(t)=e^{{\rm i}\gamma_m(t)}c_m(0)$, where $\gamma_m(t)\equiv {\rm i} \int_0^t dt'\langle m(t')|\big[\frac{\partial}{\partial t'}|m(t')\rangle\big]$ is known as the geometrical phase or Berry's phase when the evolution is cyclic.
Therefore, $|\psi(t)\rangle ={\sum}_n c_n(0) e^{{\rm i}\gamma_n(t)}e^{{\rm i}\theta_n(t)}|n(t)\rangle$. If the system starts in an eigenstate $|n\rangle$ of $\hat H_0(t=0)$, then it continues in the eigenstate $|n(t)\rangle$ of $\hat H_0(t)$, since $c_i(0)=0$ unless $i=n$, in that case $c_n(0)=1$. The solution in the adiabatic approximation is $|\psi_n(t)\rangle=e^{{\rm i}\gamma_n(t)}e^{{\rm i}\theta_n(t)}|n(t)\rangle$.

%

\end{document}